\documentclass[runningheads]{llncs}

\usepackage{amsmath}
\usepackage{hyperref}
\usepackage{cleveref}
\usepackage{graphicx}
\usepackage{tabularx, threeparttable, booktabs}

\newcommand{\bheading}[1]{{{\textbf{#1.}}}}

\newcommand{\toppone}{\texttt{top\_p=1}}
\newcommand{\llamacpp}{\texttt{llama.cpp}}

\usepackage[disable]{todonotes} %

\author{}
\date{}

\usepackage{pgfplots}
\usepackage{xcolor}
\pgfplotsset{compat=1.18}

\definecolor{cadetblue}{rgb}{0.2667,0.6667,0.6000}
\definecolor{olivedrab}{rgb}{0.6000,0.6000,0.2000}
\definecolor{skyblue}{rgb}{0.5333,0.8000,0.9333}

\newcommand{\customOrcidID}[1]{{\href{https://orcid.org/#1}{\protect\includegraphics[height=.7em]{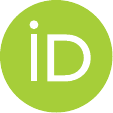}}}}

\begin{document}

\title{
Steganography Without Modification: \\ Hidden Communication via LLM Seeds
}
\titlerunning{Hidden Communication via LLM Seeds}

\author{Felix Mächtle\inst{1}\,\customOrcidID{0009-0009-2431-0322} \and
Jonas Sander\inst{1}\,\customOrcidID{0009-0007-9402-823X} \and
Sebastian Berndt\inst{2}\,\customOrcidID{0000-0003-4177-8081} \and
Ben Weimar\inst{1}\,\customOrcidID{0009-0005-2307-9887} \and 
Nils Loose\inst{1}\,\customOrcidID{0009-0003-6243-1623} \and
Thomas Eisenbarth\inst{1}\,\customOrcidID{0000-0003-1116-6973}}

\institute{Institute for IT Security, University of Lübeck, Lübeck, Germany\\
\email{\{f.maechtle,j.sander,n.loose,thomas.eisenbarth\}@uni-luebeck.de} \and
Technische Hochschule Lübeck, Lübeck, Germany\\
\email{sebastian.berndt@th-luebeck.de}}

\authorrunning{F. Mächtle \textit{et al.}}

\maketitle

\begin{abstract}
We demonstrate that widely deployed Large Language Model (LLM) inference stacks harbor a  steganographic channel that requires no modification to model weights, sampling code, or output distributions. The channel exploits a structural property of deterministic decoding: pseudo-random number generators (PRNGs) used in inverse-transform sampling produce a seed-dependent sequence of token-level probability intervals that can be reconstructed from the generated text alone. A sender encodes a secret message in the PRNG seed before generation; a receiver reconstructs the intervals and recovers the seed, and thus the hidden payload, by exhaustive search over the seed space.

We formalize two operational modes. In the \emph{known-prompt} setting, sender and receiver share the prompt, enabling exact interval reconstruction and perfect seed recovery via forced alignment. In the \emph{unknown-prompt} setting, only the generated text is available; approximate interval reconstruction combined with a maximum-hit-count scoring strategy still permits reliable recovery from sufficiently long outputs. 

Extensive experiments across six model families and five heterogeneous text domains show that, in the known-prompt setting, full 32-bit seed recovery from the complete $2^{32}$ candidate space achieves up to 100\% accuracy, depending on model and text domain, within 300~tokens and under 35~seconds on a single GPU. In the unknown-prompt setting, recovery reaches near-perfect accuracy at 600--800~tokens in about 12~seconds. We further analyze the influence of prompting strategies, tokenization ambiguities, and sampling hyperparameters on channel reliability. Moreover, we discuss several applications of our results: First, it allows for the steganographic transmission of 32 bits, but also shows that ignorance of the prompt is not a valid security assumption.

\end{abstract}

\section{Introduction}

\begin{figure}[t]
  \centering
  \includegraphics[width=0.75\columnwidth]{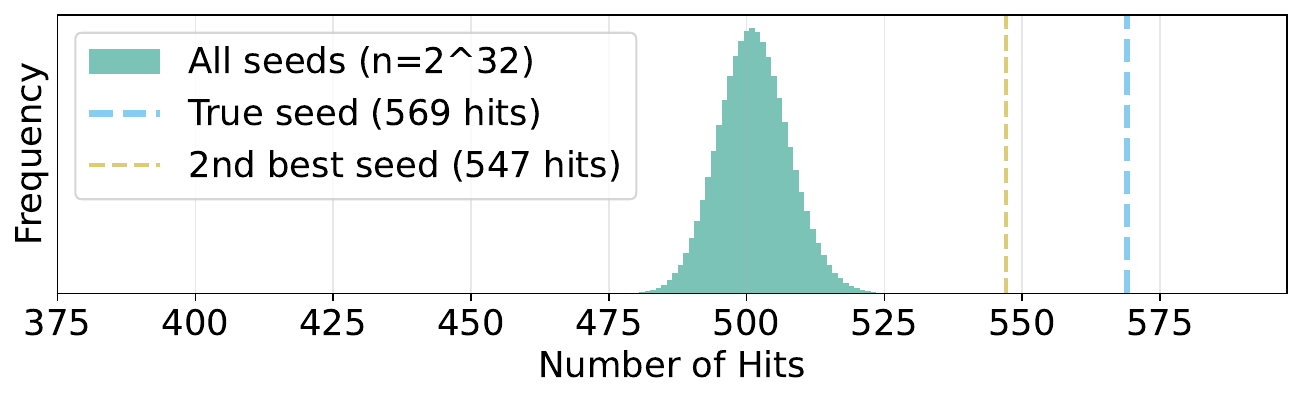}
  \caption{
From the response text alone, we exhaustively test all $2^{32}$ seeds  in $\sim$12s and count how often their simulated draws match the observed text. Nearly all seeds cluster at chance level, while a single seed stands out as a clear outlier, revealing the embedded message.}
  \label{fig:seed-hist-example}
\end{figure}

Steganography, the practice of concealing secret messages within innocuous cover media, has a long history in images, audio, and network protocols. 
A long standing problem was the creation of sufficiently many non-suspicious cover media.
With the proliferation of Large Language Models (LLMs) as a dominant medium for text generation for public websites~\cite{DBLP:journals/corr/abs-2504-08755,law2025aiwebcontent}, natural-language outputs have become an attractive new cover medium. Recent proposals embed hidden information by modifying the sampling procedure \cite{DBLP:conf/ccs/KaptchukJ0R21,DBLP:conf/emnlp/ZieglerDR19} or the model distribution itself~\cite{DBLP:journals/corr/abs-2106-02011}. Such approaches, however, require the sender to alter the decoding algorithm, which may be detectable, impractical, or undesirable.
A common problem with these new proposals is the fact that the outputs of complex LLMs follow a complicated distribution, which typically is not easy to describe mathematically due to the high non-linearity and high parameter count of today's models.
Hence, the construction of non-detectable modifications is highly complex, as discussed, e.g., in \cite{DBLP:journals/corr/abs-2210-14889,DBLP:conf/ccs/KaptchukJ0R21,DBLP:journals/iandc/BerndtL20}.

In this work, we show that no such modification is necessary. Widely deployed LLM inference stacks~\cite{llamacpp2025,vllm2025} already harbor a steganographic channel, arising from a structural property of deterministic decoding that has so far been overlooked. Modern token samplers rely on pseudo-random number generators (PRNGs)~\cite{DBLP:journals/iacr/AlichEHLMPRST25} to select the next token from the model's predicted probability distribution~\cite{DBLP:conf/nips/VaswaniSPUJGKP17}. Because PRNGs are deterministic given their seed, the entire sequence of sampled tokens is a deterministic function of the seed. We demonstrate that this determinism can be exploited for covert communication: a sender encodes a secret payload in the PRNG seed, generates seemingly ordinary text, and a receiver recovers the seed, and thus the payload, from the public output alone.

The key technical insight is that each generated token occupies a contiguous interval of the PRNG's output range, determined by the cumulative probability distribution at that position. By reconstructing these token-level intervals and testing candidate seeds against the observed sequence, the receiver can identify the unique seed whose PRNG stream is consistent with the text. The true seed emerges as a strong statistical outlier in the distribution of all seeds (\Cref{fig:seed-hist-example}). Crucially, this procedure requires neither access to internal model states nor modification of the generation process; the channel is inherent in the deterministic structure of standard sampling.

We distinguish two operational modes of the channel, differing in the information shared between sender and receiver.

\paragraph{(i) Known-prompt channel.}
If sender and receiver share the prompt and decoding configuration, the PRNG seed becomes a subliminal communication primitive. The sender encodes secret bits directly into the seed before generation. The resulting text appears natural and statistically unmodified, yet the receiver can replay the exact generation context, reconstruct the probability intervals, and recover the seed from the output. Because no token distribution is altered and no watermark is embedded, the channel leaves no statistical footprint beyond the inherent determinism of sampling.

\paragraph{(ii) Unknown-prompt channel.}
Even when the original prompt is not shared, covert communication remains possible. The receiver can approximate the probability of individual response tokens using a neutral proxy prompt and recover the seed. Although individual probability estimates are less precise, the discriminative information accumulates across tokens, and the true seed remains identifiable from sufficiently long outputs. This mode relaxes the setup requirements at the cost of requiring longer cover texts.

Extensive experiments across six model families and five heterogeneous text domains demonstrate that the channel is both reliable and practical. In the known-prompt setting, 32-bit seed recovery achieves up to 100\% accuracy,  depending on model and text domain, from 300~tokens in approximately 34~seconds on a single GPU. In the unknown-prompt setting, recovery reaches near-perfect accuracy at 600--800~tokens in about 12 seconds. We further analyze the influence of prompting strategies, tokenization ambiguities, and sampling hyperparameters on channel capacity and reliability.
Our complete implementation and dataset is publicly available on GitHub: \url{https://github.com/UzL-ITS/Hidden-Communication-via-LLM-Seeds}.

\bheading{Contributions}
This paper makes the following contributions:

\begin{itemize}
    \item We identify deterministic PRNG-based decoding as a steganographic channel in standard LLM inference stacks that requires no modification to model weights, sampling algorithms, or output distributions.
    \item We present a seed-recovery method that extracts the hidden payload from generated text alone, operating in both known-prompt and unknown-prompt settings.

    \item We empirically validate the channel across six model architectures and five text domains, and analyze the effects of sequence length, temperature, top-$k$ filtering, and prompting strategy on recovery reliability.
\end{itemize}

\paragraph{Applications.}

Our findings reveal that generation-layer randomness is not merely an implementation detail but a covert communication resource. Deterministic decoding can transform any standard LLM deployment into a steganographic system. We argue that secure deployment of generative models must therefore consider not only model weights and outputs, but also the structure of randomness consumption during decoding.
Furthermore, our findings show a surprising close relation between the PRNG seeds and the prompts. 
One important consequence of this fact is that "unknown prompts do not help", i.e., the security of a steganographic system can not solely rely on keeping the prompt unknown to the adversary.

While the presented covert channel only has a very limited bandwidth, one can increase the bandwidth by allowing multiple interactions. 
But even for a single interaction leading to the transmission of 32 bit, this transmitted information can already be very useful, e.g., by bootstrapping larger covert channel, allowing for low-rate authentication, encoding pointers into public datasets, or choosing among a limited number of operations or states~\cite{DBLP:journals/pieee/PetitcolasAK99,DBLP:conf/ih/PevnyK14,wuSecurityorientedSteganographicPayload2024,DBLP:journals/sensors/WojtunP21}.

\section{Preliminaries}

\subsection{Large Language Models (LLMs)}

Large Language Models (LLMs) are autoregressive neural networks that generate text token by token~\cite{DBLP:conf/nips/VaswaniSPUJGKP17}. Given an input sequence (prompt), the model is used iteratively to predict the next token. In modern deployments, inputs are typically structured using a \emph{chat template}, a standardized formatting scheme that encodes roles such as \emph{user} (prompt) and \emph{assistant} (response) into a single model-specific input string~\cite{huggingface_transformers_docs2025}. Conditioned on this formatted context, the model produces a probability distribution over the vocabulary, from which the next token is selected via random sampling.

\subsection{Token Sampling}

Most inference servers apply a \emph{sampling chain} that filters the raw distribution before selecting the next token. Two common filters are:

\begin{itemize}
    \item \textbf{Top-k Sampling}: Retains only the $k$ highest-probability tokens.
    \item \textbf{Nucleus (Top-p) Sampling}~\cite{DBLP:conf/iclr/HoltzmanBDFC20/NucleusSampling}: Retains the smallest set whose cumulative probability exceeds~$p$.
\end{itemize}

After filtering, the remaining tokens define a categorical distribution over the candidate set, with probabilities renormalized to sum to~1. A token is then sampled using a PRNG. Two dominant strategies exist:

\begin{itemize}
    \item \textbf{Inverse Transform Sampling}: The cumulative probabilities partition $[0,1)$ into segments proportional to the token probabilities. A single PRNG draw $u \sim \mathcal{U}(0,1)$\footnote{To be precise; pseudorandomly uniform over the interval with discrete support.} selects the token whose segment contains~$u$. Used by \llamacpp{}~\cite{llamacpp2025} and HuggingFace Transformers~\cite{huggingface_transformers_docs2025} with a Mersenne Twister PRNG.
    \item \textbf{Exponential Trick}: For each candidate token~$j$ with probability $p_j$, an independent exponential variate $q_j \sim \text{Exp}(1)$ is drawn, and the token $\arg\max_j (p_j / q_j)$ is selected. Used by vLLM~\cite{vllm2025} with Philox~\cite{DBLP:conf/sc/SalmonMDS11}, a  counter-based PRNG.
\end{itemize}

\section{Methodology}

This section describes the communication settings and the procedure for reconstructing the sampling intervals, as well as recovering the hidden seed and the datasets we use.

\subsection{Communication Model}

We consider a covert communication setting in which a \emph{sender} encodes a secret message into the PRNG seed of an LLM generation, and a \emph{receiver} recovers the message from the generated text. We assume the following conditions hold.

\begin{itemize}

    \item \textbf{Complete Output Access.}
    The receiver observes the generated text exactly as produced, without paraphrasing or truncation.

    \item \textbf{Deterministic Seeded Sampling.}
    The model uses a deterministic PRNG for token sampling, whose implementation is known to both parties. This assumption holds for common LLM inference frameworks such as \llamacpp{}.

    \item \textbf{Sampling Parameters.} We assume that both sender and receiver know the sampling parameters (e.g.,\ by using the defaults).

    \item \textbf{Model Access.} 
The sender must be able to set the PRNG seed. 
The receiver must have access to the same model and be able to obtain the post-filtering token probabilities, either through local model access or via an API that exposes output probabilities, i.e., a completion API as offered by many providers.

    \item \textbf{Shared Prompt or  \toppone{}.}
    The sender and receiver may share the exact prompt $P$ in advance. As an alternative, we require \toppone{}, which is natively satisfied by the default values of, e.g.,  OpenAI~\cite{microsoft_azure_openai_api_2026}, Grok~\cite{xai_openapi_2026}, or Mixtral~\cite{mistral_api_2026}, but can be configured for any model.

    \item \textbf{Seed Space / Message Space.}
The PRNG seed space is finite (e.g., 32-bit). Recovery relies on exhaustive search, so the seed space must be tractable for brute-force enumeration. In practice, $2^{32}$ seeds are searchable with GPU parallelization. Thus, the payload size must be chosen accordingly.
\end{itemize}

The above assumptions are not hypothetical but reflect standard configurations of widely used inference stacks.
For example, running an open-weight model from HuggingFace via \llamacpp{} with default decoding settings already satisfies all required properties.
In such setups, the sender can embed a message in the seed without modifying model weights or sampling code.

\bheading{Threat Model}
In this work, we only consider passive attackers that eavesdrop the communication between the sender and the receiver, but do not have the capability to actively modify the communication. 
Furthermore, we assume that the message to be sent is uniformly distributed, which can be achieved, e.g., by using encryption schemes with pseudorandom ciphertexts.

\subsection{Ambiguous Tokenization}
\label{sec:main:Ambiguous-Tokenization}
\label{sec:main:forced-alignment}

Because the receiver only receives the final response text from the sender, he must first convert it into the original tokens. However, encoding and decoding in modern LLM tokenizers do not form a strict bijection: the same surface text can correspond to multiple valid token sequences. We call this \emph{ambiguous tokenization}. During autoregressive generation, tokens are selected one at a time conditioned on context. Re-tokenizing the decoded output with a tokenizer may segment it differently, e.g., merging two short tokens into one (e.g. 'Hell' + 'o' becoming 'Hello'). This might cause mismatches between the original and the reconstructed tokens.

We address this challenge with two strategies, depending on whether the original prompt is available:

\bheading{Known-Prompt}
When the prompt is known, we resolve tokenization ambiguities via \emph{forced alignment with proactive branching}. We replay the generation context and consume the output text position by position. At each step, we call the LLM to obtain the filtered token probability distribution and scan it for all tokens whose byte representation forms a valid prefix of the remaining text. If multiple matching tokens exist with different byte lengths, the text at this position is ambiguous: the model may have originally produced any of them.

We handle this by branching: the default tokenization (as produced by the standard tokenizer on the remaining text) is followed as the primary path, while each alternative token spawns an independent branch that duplicates the current search state and continues with its own token fed into the LLM context.

Seed elimination proceeds interleaved with alignment. We maintain a bitmap of $2^{32}$ candidate seeds on the GPU. As ranges are collected along each path, they are flushed to the GPU to eliminate non-matching seeds. Branches whose bitmap reaches zero survivors are pruned immediately. This early pruning keeps the number of active branches small in practice: the vast majority of branches are eliminated within a few tokens, as  the correct tokenization path  dominates.

\bheading{Unknown-Prompt}
\label{sec:main:offset-matching}
When the original prompt is unknown, forced alignment is not feasible: the generation context cannot be reconstructed, and tokens with zero probability may occur naturally rather than indicating a mismatch. We therefore decode the text into tokens directly, accepting that tokenization mismatches may occur. 

To mitigate their impact, later during seed recovery (see \Cref{sec:main:seed-recovery}) we do not evaluate a single token position only. Instead, we simultaneously consider three possible alignments ($i{-}1$, $i$, and $i{+}1$) when comparing PRNG outputs to the reconstructed intervals. This allows small tokenization shifts to be tolerated without requiring knowledge of the exact mismatch location.

\subsection{Token Probability Range Extraction}

In order to reconstruct the seed, the receiver must first reconstruct the probability intervals from which each token in the generated text was sampled with high precision. Given an observed sequence of tokens $t_1, t_2, \dots, t_n$, the receiver reproduces the sampling process by running the identical LLM on either the shared prompt or a reconstructed prompt and computing the next-token probability distribution at each position.

For each position $i$, the receiver applies the same filtering strategy (e.g., top-$k$ or nucleus top-$p$) and temperature as the sender's generation configuration. This yields a reduced candidate set of tokens $C_i = \{c_{i,1}, c_{i,2}, \dots\}$ with associated normalized probabilities $p_{i,1}, p_{i,2}, \dots$. These probabilities define a set of disjoint cumulative probability intervals that partition the unit interval $[0,1)$. 
We store the integer representation of these intervals according to the precision of the PRNG in use (e.g., an interval of $0-0.5$ becomes $0 - \frac{2^{32}}{2}$). This allows direct comparison between simulated PRNG outputs and the intervals derived from reconstructed probabilities, making the computation more efficient, as it operates on integers rather than floating-point values.

\subsection{Seed Recovery}
\label{sec:main:seed-recovery}

Given the extracted probability intervals, the receiver searches for the PRNG seed, i.e.,  the embedded message, that produced the observed token sequence. Let $[l_i, h_i) \subset [0, 2^{32})$ denote the integer-scaled interval for token position~$i$, and let $u_i(s)$ denote the integer PRNG output at position~$i$ under candidate seed~$s$. We define the \emph{hit count} of a seed as
\[
H(s) \;=\; \sum_{i=1}^{n} \mathbf{1}\bigl[u_i(s) \in [l_i, h_i)\bigr],
\]
where the sum runs over all token positions.

\bheading{Known-Prompt} 
When the prompt is known, the reconstructed intervals are exact. The true seed therefore produces a hit at every position. Seed recovery succeeds if and only if exactly one seed satisfies $H(s) = n$, i.e., the seed matches all reconstructed intervals.

\bheading{Unknown-Prompt}
When the prompt is unknown, the reconstructed intervals are only approximate because the proxy prompt induces different logits than the original. Even the true seed will therefore not match at every position, making perfect matching infeasible.

To tolerate tokenization shifts (see \Cref{sec:main:offset-matching}), we evaluate multiple alignment offsets. For each candidate seed $s$ and offset $\delta \in \{-1,0,1\}$, we compute
\[
H_{\delta}(s) \;=\; \sum_{i=1}^{n} \mathbf{1}\bigl[u_{i+\delta}(s) \in [l_i, h_i)\bigr].
\]
The final score for a seed is the maximum hit count across offsets,
\[
H^*(s) = \max_{\delta \in \{-1,0,1\}} H_{\delta}(s),
\]
and the recovered seed is
\[
s^* = \arg\max_{s \in \{0,\dots,2^{32}\}} H^*(s).
\]

The true seed is expected to be a statistical outlier in the resulting score distribution (see \Cref{fig:seed-hist-example}). For a random (incorrect) seed, each position contributes a match with probability equal to the interval's fractional width. As  hits across positions are  independent, the background score is a sum of independent Bernoulli variables, so it has a Poisson binomial distribution and concentrates around its mean. The true seed, by contrast, benefits from correlated matches at positions where the reconstructed interval overlaps with the original, yielding a score that separates clearly from the background (see \Cref{fig:seed-hist-example} as an example).

\bheading{Applicability to exponential-trick sampler}
\label{sec:main:vllm}
While not directly applicable, the methodology also extends to exponential-trick samplers as used in, e.g., vLLM~\cite{vllm2025}. Because each token draws one exponential variate per candidate (after filtering), the observed token is necessarily the one that maximized $p_j / q_j$ under the true seed's PRNG state. We can therefore replace the interval-based hit test with a simulation: for each candidate seed, recompute $\arg\max_j (p_j / q_j)$ at every position and count matches with the observed sequence. The true seed produces a statistically dominant hit count. While we verified this experimentally, the remainder of this paper focuses on \llamacpp{}'s inverse-transform sampler: it requires no GPU for text generation, runs on commodity hardware, and is thus more accessible to practitioners.

\subsection{Constant-Draw Assumption}
\label{sec:main:topp-less-than-one}
In the unknown-prompt mode, our seed-recovery pipeline requires each token to consume a constant number of PRNG variates, the \emph{Constant-Draw Assumption (CDA)}. With inverse-transform sampling and \toppone{}, CDA holds trivially: exactly one draw per token.

When \texttt{top\_p<1}, CDA is typically violated. After nucleus filtering, it may occur that a single token already exceeds the cumulative probability threshold~$p$. In such cases (e.g., in the sentence “The 44th U.S. president was Barack ”, where the next token is almost certainly “Obama”), implementations such as \llamacpp{} bypass sampling and deterministically return that token without querying the PRNG. We call these \emph{no-draw tokens (NDTs)}. NDTs introduce an unknown, input-dependent offset between token index and PRNG index, collapsing the discriminative power of hit counting. Conversely, if the prompt is shared, the receiver can detect NDTs and skip the corresponding PRNG advances.

The fundamental issue is thus not \texttt{top\_p} per se but variable PRNG consumption. We adopt \toppone{} as a practical sufficient condition for CDA: every position consumes exactly one uniform variate, enabling a stable mapping from seed to token index.

\begin{table}[t]
    \centering
    \caption{Example prompts for each dataset.}
    \begin{tabularx}{\columnwidth}{lX}
\toprule
Dataset & Prompt \\ \midrule

PubMed Q\&A & Do mitochondria play a role in remodelling lace plant leaves during programmed cell death? \\
Reddit WP & You are a creative writing assistant. Use the provided prompt to create a coherent and well-developed story. Ensure the plot follows the prompt's guidelines, with a clear logical progression, substantial narrative development, and fluent, polished language. Every person in the world undergoes a `` goodness '' test . It 's designed to give a score from 1 to 200 , where 1 is pure evil , and 200 is an angel in human body . Then the world is divided into 200 zones , where people can live among their own kind .\\
XSUM EN & Please write a news story about the following topic: Clean-up operations are continuing across the Scottish Borders and Dumfries and Galloway after flooding caused by Storm Frank.\\
XSUM DE & Please write a news story about the following topic: Clean-up operations are continuing across the Scottish Borders and Dumfries and Galloway after flooding caused by Storm Frank.  Write in German! \\

         Code Contest & I want you to act as a software developer. I will provide you with some requirements about
a software, and it will be your job to implement the functionality in Python. Do not write
explanations, just reply with the code. My request is: Problem description.
Vipul is a hardworking super-hero who maintains the bracket ratio of all the strings in the world. Recently he indulged himself in saving the string ... 
\\ \bottomrule

    \end{tabularx}
    \label{tab:appenidx:prompt-examples}
\end{table}

\subsection{Prompt Dataset}
\label{sec:main:dataset}

We curate a multi-domain prompt suite covering five generation use cases, following~\cite{DBLP:conf/iclr/BaoZTY024} but extending to programming code. The domains span (i)~program synthesis, (ii)~technical prose, (iii)~open-ended narrative, and (iv)~factual news in two languages. Example prompts are listed in Table~\ref{tab:appenidx:prompt-examples}.

\bheading{Code Contest}
We draw problem statements from the DeepMind Code Contest dataset~\cite{DBLP:journals/corr/abs-2203-07814/DeepMindCodeContestDataset}, each prefixed with a brief instruction to produce a Python solution following Xu~\emph{et~al.}~\cite{DBLP:journals/tosem/XuNGLWLY25}.

\bheading{PubMed Q\&A}
We use questions from PubMed QA~\cite{DBLP:conf/emnlp/JinDLCL19/PubMedQA} as prompts, without providing abstracts or gold answers.

\bheading{Reddit WP}
We use the Reddit Writing Prompts dataset~\cite{DBLP:conf/acl/LewisDF18/RedditWritingPrompts}, prepending a minimal instruction template following Keigo~\cite{akeigo-llama3.1-storygeneration} to elicit long-form narratives without stylistic priming or length targets.

\bheading{XSUM EN}
From the XSUM corpus~\cite{DBLP:conf/emnlp/NarayanCL18/XSUM}, we convert each item into a prompt-based task: we prepend \emph{``Please write a news story about the following topic:''} and provide the gold summary as the topic. Unlike prior work~\cite{DBLP:conf/iclr/BaoZTY024}, which uses prefix completion, this formulation elicits full prompt-driven generation.

\bheading{XSUM DE}
To evaluate cross-lingual transfer, we reuse the XSUM~EN prompts and append an instruction to write in German.

To characterize the resulting text distributions, \Cref{tab:prompt-dataset-stats} reports average statistics over 250 prompts (50 per domain) and all six models, generated with each model's default parameters but \toppone{}. The table summarizes answer lengths, the token ambiguity, defined as the Levenshtein distance between the original and reconstructed token-ID sequences (see \Cref{sec:main:Ambiguous-Tokenization}), and the proportion of \emph{no-draw} tokens (see \Cref{sec:main:topp-less-than-one}). Under \toppone{}, the PRNG is still advanced at these positions, so ``no-draw'' is a misnomer. However, we retain the term because a single token holds all the probability mass.

\begin{table}[t]
    \centering
    \caption{Dataset statistics (averaged across models).}
    \label{tab:dataset_stats_main}
    \label{tab:prompt-dataset-stats}

\begin{tabularx}{0.9\columnwidth}{Xccc}
\toprule
Dataset & Output Length & Token Ambiguity & No-Draw Tokens \\
 & (tokens) & (edit distance) & (\%) \\
\midrule
PubMed Q\&A & 635.4 $\pm$ 308.4 & 3.2 $\pm$ 10.6 & 8.1 $\pm$ 7.2 \\
Reddit WP & 749.6 $\pm$ 333.3 & 1.2 $\pm$ 2.8 & 3.9 $\pm$ 6.8 \\
XSUM EN & 646.5 $\pm$ 218.8 & 2.0 $\pm$ 4.1 & 5.2 $\pm$ 4.8 \\
XSUM DE & 757.6 $\pm$ 231.6 & 3.8 $\pm$ 6.1 & 6.7 $\pm$ 6.9 \\
Code Contest & 574.1 $\pm$ 376.9 & 3.0 $\pm$ 9.7 & 24.4 $\pm$ 23.8 \\

\bottomrule
\end{tabularx}

\end{table}

\subsection{Implementation}
\label{sec:main:implementation}

All experiments use \llamacpp{}~\cite{llamacpp2025}. We specifically target \llamacpp{}, as it enables running LLMs on hardware without a GPU. This makes the setup easily accessible to senders without requiring expensive GPU hardware. To simulate the receiver, we add a read-only hook to \llamacpp{} that captures the post-filtering probability vector after each sampling step, enabling exact interval reconstruction without modifying decoding. Notably, this modification is only required for the receiver.
For efficient seed recovery, we wrote custom CUDA kernels for the respective experiments.

\section{Experiments}

We evaluate the reliability and practical constraints of seed recovery across models, domains, and decoding settings in both known- and unknown-prompt scenarios. All experiments use six instruction-tuned models spanning four architecture families and a range of parameter counts: Gemma~3 4B\footnote{\url{https://huggingface.co/ggml-org/gemma-3-4b-it-GGUF}}, Qwen~2.5 3B\footnote{\url{https://huggingface.co/Qwen/Qwen2.5-3B-Instruct-GGUF}}, Llama~3.2 3B\footnote{\url{https://huggingface.co/unsloth/Llama-3.2-3B-Instruct-GGUF}}, Phi-4 Mini\footnote{\url{https://huggingface.co/unsloth/Phi-4-mini-instruct-GGUF}}, Qwen3 4B\footnote{\url{https://huggingface.co/unsloth/Qwen3-4B-Instruct-2507-GGUF}}, and Qwen3 30B-A3B (MoE)\footnote{\url{https://huggingface.co/unsloth/Qwen3-30B-A3B-Instruct-2507-GGUF}}. All models are run in Q4\_K\_M quantization via \llamacpp{}. For the unknown-prompt experiments (RQ2--RQ4), text generation ran on ARM while interval reconstruction and seed recovery ran independently on x86, validating that the channel is robust to cross-platform floating-point differences.
We address the following research questions:
\begin{itemize}
    \item[]\textbf{RQ1:} How accurately can we recover the seed when the prompt is known?
    \item[]\textbf{RQ2:} How do prompting strategies influence the quality of range reconstruction?
    \item[]\textbf{RQ3:} How do hyperparameters, i.e., top-$k$ and temperature, influence seed recovery?
    \item[]\textbf{RQ4:} How effective is seed recovery without knowledge of the prompt?
\end{itemize}

\subsection{Known-Prompt Seed Recovery}

We first evaluate the known-prompt setting, which employs forced alignment (\Cref{sec:main:forced-alignment}) and does \emph{not} require \toppone{}. The goal is to determine the minimum sequence length at which the true 32-bit seed can be uniquely recovered from the full $2^{32}$ candidate space.

\bheading{Experimental Design}
For each of the six models, we generated 125~samples (25~prompts per dataset domain) using each model's default sampling configuration. Each sample was assigned a distinct random 32-bit seed drawn uniformly from $[0, 2^{32})$, and generation was capped at 500~tokens.
We applied forced alignment with proactive branching (\Cref{sec:main:forced-alignment}) to extract ranges using the known prompt. A per-sample timeout of 100~seconds was enforced. A trial succeeds if exactly one seed survives and equals the true seed.

\bheading{Results}
\begin{figure}[t]
\centering
\includegraphics[width=\linewidth]{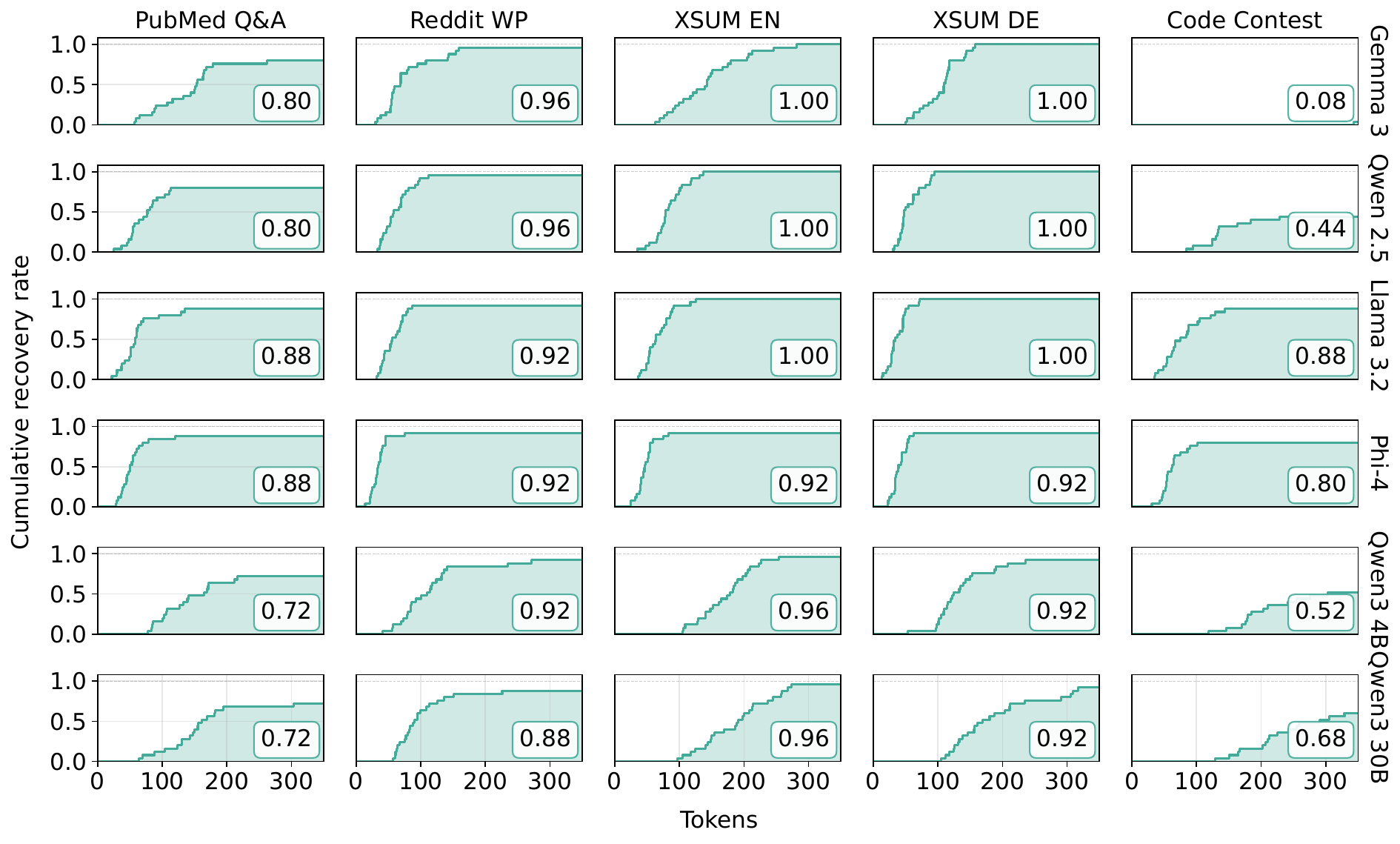}
\caption{Known-prompt seed recovery (exhaustive $2^{32}$ bitmap search with proactive branching). Each cell shows the cumulative fraction of samples for which the seed is correctly recovered. The label in each cell reports the final accuracy. Columns correspond to dataset domains, rows to models.}
\label{fig:known_prompt_results}
\end{figure}
\Cref{fig:known_prompt_results} presents seed recovery accuracy across all six models and test domains. Overall, recovery succeeds for 84.5\% of samples. Llama~3.2 achieves the highest rate (93.6\%), followed by Phi-4 (88.8\%) and Qwen~2.5 (84.0\%), while Qwen3~4B and Qwen3~30B reach 80.8\% and 83.2\%, respectively. Gemma~3 is the lowest at 76.8\%.

News summarization domains achieve near-perfect recovery: XSUM~EN (97.3\%) and XSUM~DE (96.0\%), followed by Reddit~WP (92.7\%) and PubMed~Q\&A (80.0\%). Code Contest is the hardest domain  (56.7\%). Code samples produce more no-draw tokens because keywords, brackets, and indentation yield near-deterministic distributions with probability close to~1, reducing discriminative power (see  \Cref{tab:prompt-dataset-stats}).

Failures fall into two categories: \emph{timeouts} (44 out of 116 failures), where excessive branching from tokenization ambiguity exhausts the 100-second budget, and \emph{exhausted} branches (72 out of 116), where all branches are explored but none yields a unique seed, typically because too few informative ranges exist (as in code samples). On successful samples, the average solving time is 34~seconds on a single NVIDIA H200.

\bheading{Discussion}
These results demonstrate that, in the known-prompt setting, seed recovery is both feasible and reliable across diverse model families. The 84.5\% overall accuracy, and 92--97\% on news and creative-writing domains, confirms that the per-token discriminative power of probability intervals is sufficient to uniquely recover a 32-bit embedded message from a few hundred tokens.

The domain-level differences are well explained by the fraction of \emph{no-draw} tokens reported in \Cref{tab:prompt-dataset-stats}. Code Contest exhibits the highest no-draw rate ($24.4\%$), roughly five times that of Reddit~WP ($3.9\%$) and four times that of XSUM~EN ($5.2\%$). This directly accounts for Code Contest's lower recovery rate: fewer informative tokens means more text is needed to accumulate sufficient evidence to uniquely identify the seed. 

The primary remaining obstacles are therefore (1)~low-entropy domains such as code, where no-draw tokens dominate and too few positions carry discriminative information, and (2)~tokenization ambiguity, which can cause excessive branching and timeouts.

\subsection{Prompting Strategy Impact on Range Reconstruction}

\begin{table}[t]
\centering
\caption{Average range overlap by task (across all models).}
\label{tab:prompt_overlap_by_task}
\begin{tabularx}{\linewidth}{Xcccc}
\toprule
Task & No Prompt & Empty Prompt & Continue Prompt & Generated Prompt \\
\midrule
PubMed Q\&A & 0.64 $\pm$ 0.12 & 0.71 $\pm$ 0.10 & 0.70 $\pm$ 0.09 & 0.70 $\pm$ 0.09 \\
Reddit WP & 0.52 $\pm$ 0.12 & 0.60 $\pm$ 0.09 & 0.60 $\pm$ 0.09 & 0.57 $\pm$ 0.08 \\
XSUM DE & 0.63 $\pm$ 0.08 & 0.68 $\pm$ 0.09 & 0.69 $\pm$ 0.09 & 0.68 $\pm$ 0.08 \\
XSUM EN & 0.57 $\pm$ 0.13 & 0.65 $\pm$ 0.11 & 0.65 $\pm$ 0.11 & 0.64 $\pm$ 0.10 \\
Code Contest  & 0.74 $\pm$ 0.09 & 0.76 $\pm$ 0.10 & 0.76 $\pm$ 0.10 & 0.76 $\pm$ 0.10 \\
\midrule
\textbf{Average} & \textbf{0.62 $\pm$ 0.08} & \textbf{0.68 $\pm$ 0.06} & \textbf{0.68 $\pm$ 0.06} & \textbf{0.67 $\pm$ 0.07} \\
\bottomrule
\end{tabularx}
\end{table}

The first experiment focused on the trivial case when the sender and receiver share the same prompt. However, this might not be feasible in reality. Thus, the following experiments evaluate a more realistic scenario in which the prompt is unknown. 
When the prompt is unknown, the accuracy of our method depends heavily on how well the probability intervals can be reconstructed. This experiment evaluates four prompting strategies for interval reconstruction.

\bheading{Experimental Design}
For each sample from our dataset (see \Cref{sec:main:dataset}), we extracted the ground-truth intervals, then reconstructed them using four  alternative prefixes: \emph{nothing}, i.e., no chat template,  \emph{empty}, i.e., an empty prompt ('') in the chat template, a short \emph{generic} prompt  (\emph{Continue the following text:}), and a \emph{generated} prompt  synthesized by GPT-4o. Quality was measured by the mean normalized interval overlap ($\mathcal{O}_{\mathrm{norm}}$) between original interval $r_1$ and reconstructed interval $r_2$,
\[
\mathcal{O}_{\mathrm{norm}} \;=\; \frac{\max\!\bigl(0,\; \min(r_1^{\text{end}}, r_2^{\text{end}}) - \max(r_1^{\text{start}}, r_2^{\text{start}})\bigr)}{r_1^{\text{end}} - r_1^{\text{start}}}\!,
\] 
aggregated across tokens, models, and tasks. Results are stratified by dataset in \Cref{tab:prompt_overlap_by_task} and by model in \Cref{tab:prompt_overlap_results}.

\bheading{Results}
\begin{table}[t]
\centering
\caption{Average range overlap by prompting strategy across models.}
\label{tab:prompt_overlap_results}
\begin{tabularx}{\linewidth}{Xcccc}
\toprule
Model & No Prompt & Empty Prompt & Continue Prompt & Generated Prompt \\
\midrule
Gemma 3 & 0.55 $\pm$ 0.10 & 0.75 $\pm$ 0.06 & 0.74 $\pm$ 0.06 & 0.73 $\pm$ 0.05 \\
Qwen 2.5 & 0.58 $\pm$ 0.07 & 0.65 $\pm$ 0.06 & 0.66 $\pm$ 0.07 & 0.65 $\pm$ 0.06 \\
Llama 3.2 & 0.70 $\pm$ 0.07 & 0.66 $\pm$ 0.08 & 0.66 $\pm$ 0.08 & 0.66 $\pm$ 0.07 \\
Phi-4 & 0.45 $\pm$ 0.05 & 0.51 $\pm$ 0.05 & 0.51 $\pm$ 0.04 & 0.51 $\pm$ 0.04 \\
Qwen3 4B & 0.68 $\pm$ 0.05 & 0.73 $\pm$ 0.06 & 0.73 $\pm$ 0.06 & 0.71 $\pm$ 0.05 \\
Qwen3 30B & 0.71 $\pm$ 0.05 & 0.75 $\pm$ 0.06 & 0.75 $\pm$ 0.06 & 0.74 $\pm$ 0.04 \\
\midrule
\textbf{Average} & \textbf{0.61 $\pm$ 0.10} & \textbf{0.67 $\pm$ 0.09} & \textbf{0.67 $\pm$ 0.09} & \textbf{0.66 $\pm$ 0.08} \\
\bottomrule
\end{tabularx}
\end{table}
Dataset-wise, neutral reconstructed prompts perform as well as or slightly better than generated prompts: empty and generic prefixes both achieve $0.68\pm0.06$ average overlap, while generated prompts reach $0.67\pm0.07$. 

Model-wise (\Cref{tab:prompt_overlap_results}), Gemma~3 and the Qwen~3 variants attain the strongest overlaps. However, the relative ordering of prompting strategies is stable across architectures: additional semantic context does not improve interval alignment and can introduce mild degradation.

\bheading{Discussion}
The prompt shifts the logits before the autoregressive continuation begins, propagating misalignment through the sequence. An empty prompt perturbs the distribution less, keeping reconstructed intervals closer to the originals. 
Practically, attempts to reconstruct the missing prompt are unnecessary and can be counterproductive. A neutral or empty prefix maximizes interval overlap.

\subsection{Influence of Hyperparameters on Seed Recovery}
\label{sec:experiments:hyperparameters}

In our earlier experiments, we fixed sampling parameters to each model's defaults. In practice, however, users or deployment configurations can  alter the temperature and top-$k$ filtering. Because both parameters reshape the next-token probability distribution before sampling, they directly affect the width of the cumulative probability intervals from which seeds are recovered. This experiment systematically varies temperature and top-$k$ to quantify their influence on seed recovery accuracy in the unknown-prompt setting.

\begin{figure}[t]
    \centering
    \includegraphics[width=0.75\linewidth]{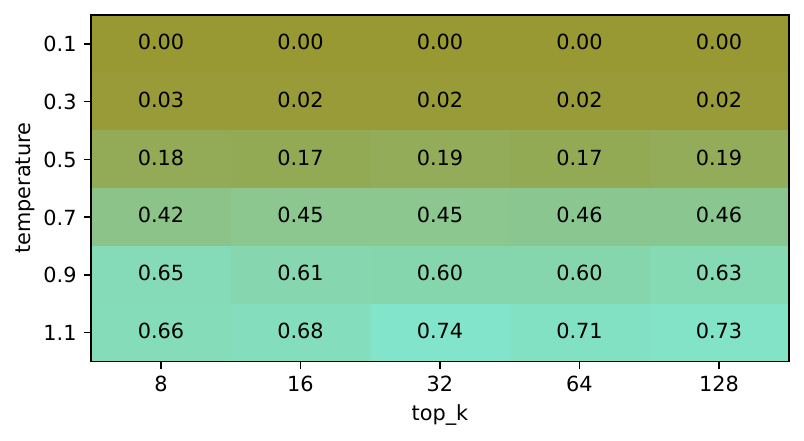}
    
    \caption{Seed recovery success rate for different temperatures and top-$k$. Temperature is the dominant factor, top-$k$ has a smaller effect.}
    \label{fig:hyperparameter_heatmap}
\end{figure}

\bheading{Experimental Design}
We use the prompt-dataset from \Cref{sec:main:dataset} and swep a grid of 30 temperature$\times$top-$k$ configurations. For each configuration, we generate ten samples per domain  with up to 1\,000 tokens with a random seed from $[0, 10\,000]$. Ranges are reconstructed with an empty prefix and seed recovery searched $10^6$ candidates. A trial succeeds if the recovered seed matches the true seed. \Cref{fig:hyperparameter_heatmap} reports success rates averaged across models and domains.

\bheading{Results}
Temperature is the decisive parameter. At $T{=}0.1$, recovery fails completely ($0.00$). Success rises monotonically: $0.02$--$0.03$ at $T{=}0.3$, $0.17$--$0.19$ at $0.5$, $0.42$--$0.46$ at $0.7$, and $0.66$--$0.74$ at $1.1$ (maximum: $0.74$ at $T{=}1.1,\, k{=}32$). Top-$k$ exerts have a smaller  influence: within any temperature row, success rates vary by at most $0.08$ across $k \in [8, 128]$.

\bheading{Discussion}
The temperature dependence has a straightforward explanation. Low temperatures concentrate probability mass on few tokens, widening their cumulative intervals toward $[0,1)$; any PRNG draw matches, so all candidate seeds achieve similar hit counts. Higher temperatures narrow the intervals, making each position more discriminative and causing the true seed's hit count to separate from the background.
Top-$k$'s smaller variance follows from the same reasoning: filtering removes  low-probability tail tokens with narrow intervals.

These findings have direct practical implications for the steganographic channel. Low-temperature deployments substantially hinder message extraction. Fortunately, many production configurations use moderate temperatures ($0.7$--$1.0$), a regime in which recovery rates already exceed 42\,\%.

\subsection{Impact of Sequence Length on Seed Recovery}
\begin{figure*}[t]
\centering
\includegraphics[width=\textwidth]{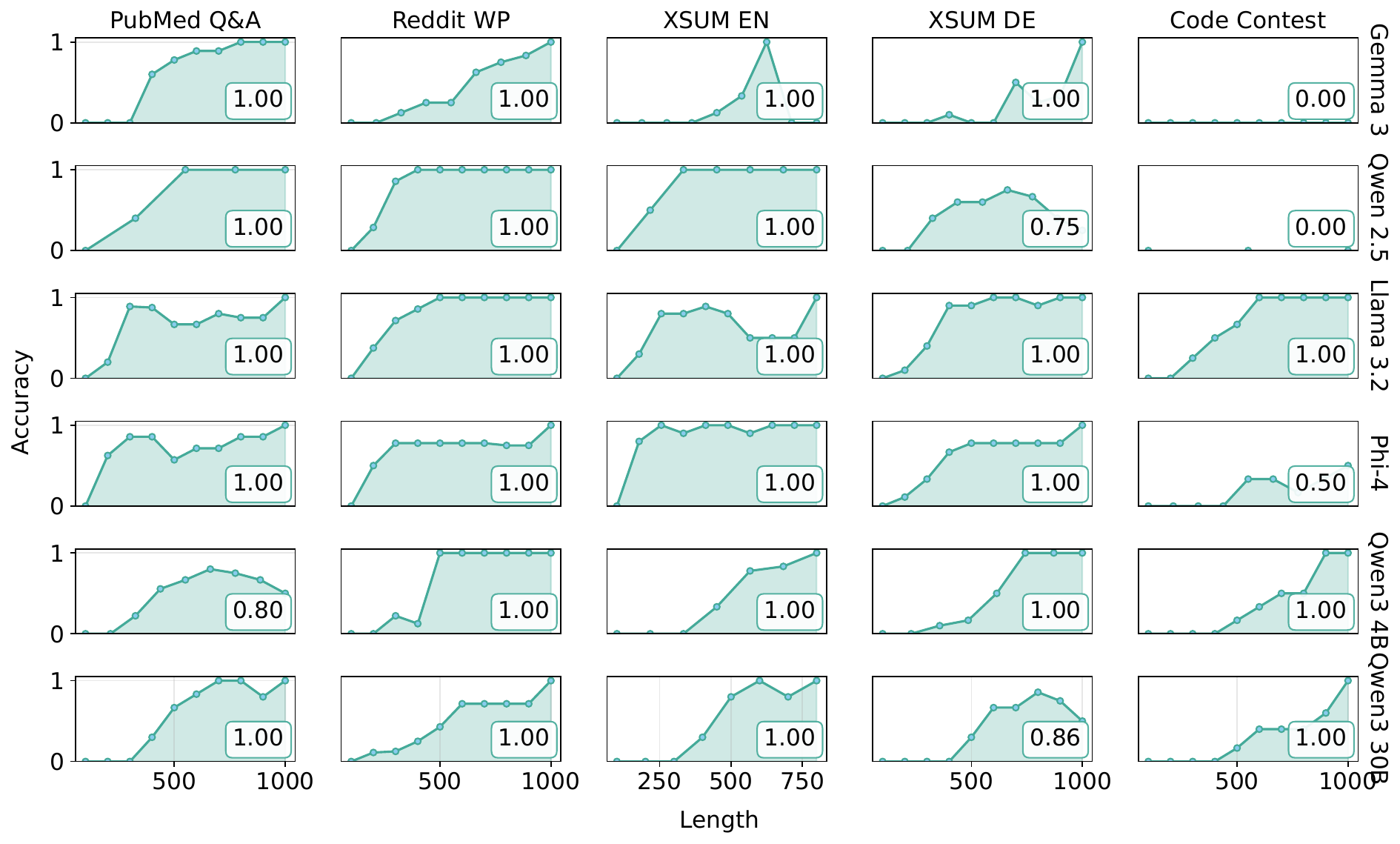}
\caption{Seed recovery accuracy for unknown prompts. Each exhaustive $2^{32}$ GPU search took around 12 seconds on a H200 GPU. The number in each subplot indicates the maximum accuracy attained independent of the token count.}
\label{fig:seed_length_accuracy}
\end{figure*}

The previous experiment identified temperature as the dominant hyperparameter for seed recovery. We now fix the sampling configuration to the best-performing setting from that experiment and study how the number of observed tokens affects recovery accuracy in the unknown-prompt mode.

\bheading{Experimental Design}
From the grid corpus, we selected the samples at $T{=}1.1$, $k{=}32$. For each, we truncated ranges to 100--1\,000 tokens in 100-token steps, discarding samples shorter than the required length, and search over the full $2^{32}$ seed space on the GPU with offset matching ($\pm1$ token). Note that reconstructing the ranges takes exactly as long as generating the text. Hence, all timings reported below refer to the seed search only. \Cref{fig:seed_length_accuracy} shows the resulting length--accuracy curves per model--task pair.

\bheading{Results}
At 100~tokens, accuracy is near zero across all pairs. However, Recovery improves rapidly with length. Most panels reach $100\%$ accuracy by 600--800~tokens, and 24 of 30 model--task cells attain perfect accuracy at their maximum observed length. Llama~3.2, Phi-4, Qwen~2.5, and Qwen3~4B reach $1.00$ across all or nearly all domains.

Departures from perfection concentrate in code generation: Gemma~3 and Qwen~2.5 remain at~$0.00$ on Code Contest, and Phi-4 plateaus at~$0.50$. Outside code, shortfalls are minor. The steepest gains occur between 300--600~tokens; Reddit~WP converges fastest. Non-monotonic dips at longer lengths reflect token ambiguity beyond the $\pm\,1$ offset.

\bheading{Discussion}
These results confirm that sequence length is the primary lever for successful message recovery once a favorable temperature regime has been established. The near-universal convergence to $1.00$ demonstrates that the per-token interval widths are narrow enough for the true seed to be uniquely identified, even in the unknown-prompt mode where only approximate interval reconstruction is possible.

The persistent difficulty with code stems from no-draw tokens (keywords, brackets, and indentation) that have low discriminative power and limit the available evidence. 
On the other hand, natural-language domains (PubMed~Q\&A, Reddit~WP) converge fastest, as nearly every position contributes discriminative information.

Practically, these findings establish a clear operational guideline: for steganographic message recovery at $T{\geq}1.0$ and \toppone{}, sequences of 600--800~tokens suffice for near-certain extraction across most model families and domains. Shorter spans remain viable for constrained domains with high per-token entropy, while code-heavy outputs require longer contexts.

\section{Related Work}

\label{sec:related-work}

Steganography, the art of concealing the existence of a message
within an innocent-looking carrier, has a long history predating
digital communication~\cite{DBLP:conf/eurocrypt/Simmons84}.

\bheading{Subliminal channels and kleptography}
The theoretical foundation for exploiting protocol randomness as a
covert channel originates with
Simmons~\cite{DBLP:conf/eurocrypt/Simmons84,DBLP:conf/eurocrypt/Simmons93}, who demonstrated
subliminal channels in digital signature schemes.
Young and Yung~\cite{DBLP:conf/crypto/YoungY96,DBLP:conf/eurocrypt/YoungY97} formalised
\emph{kleptography}, the theft of information through black-box
cryptographic implementations, showing that tampered devices can
leak secret keys through ostensibly normal public outputs.
Bellare \emph{et~al.}~\cite{DBLP:conf/crypto/BellarePR14} formalised Algorithm
Substitution Attacks (ASAs), proving that a subverted encryption
algorithm can exfiltrate plaintexts through ciphertexts
indistinguishable from honest ones.
Subsequently, Berndt and Liskiewicz~\cite{DBLP:conf/ccs/BerndtL17}  proved a 
formal equivalence between successful ASAs and secure stegosystems,
establishing that exploiting PRNG determinism in any
system, including LLM inference, for covert communication is
formally equivalent to conducting an ASA against that system.

Our work instantiates this class of attacks in a new domain: the
randomness consumed during LLM token sampling serves the same role
as the nonce in a signature scheme, providing bandwidth for a
subliminal channel that requires \emph{no substitution} of the
underlying algorithm.

\bheading{Neural text steganography}
Generative steganography embeds secret bits by controlling the
randomness consumed during text generation.
Fang \emph{et~al.}~\cite{DBLP:conf/acl/FangJA17} were among the first to use neural
language models for steganographic text generation, followed by
Yang \emph{et~al.}~\cite{DBLP:journals/tifs/YangGCHZ19}, who introduced fixed- and
variable-length coding strategies with RNNs.
Ziegler \emph{et~al.}~\cite{DBLP:conf/emnlp/ZieglerDR19} proposed arithmetic coding
over GPT-2 token distributions, treating the secret message as the
random source that drives the LLM sampling, a mechanism conceptually
identical to our covert channel, except that it needs to be deployed intentionally.
Dai \emph{et~al.}~\cite{DBLP:conf/acl/DaiC19} refined this with
patient-Huffman coding and KL-divergence constraints, while
Shen \emph{et~al.}~\cite{DBLP:conf/emnlp/ShenJH20} introduced self-adjusting arithmetic
coding that adapts the truncation threshold per token.

Three recent systems achieve provably\footnote{As usual, \emph{provably secure} means that a security proof is possible under certain complexity-theoretic assumptions.} secure steganography.
Kaptchuk \emph{et al.}~\cite{DBLP:conf/ccs/KaptchukJ0R21} (Meteor) replace
the language model's RNG with pseudorandom bits derived from an
encrypted message and adapt the encoding rate to local entropy. Our
unintentional covert channel is, in effect, an ``accidental Meteor.''
De~Witt \emph{et~al.}~\cite{DBLP:journals/corr/abs-2210-14889} proved that a steganographic
scheme is perfectly secure if and only if it is induced by a coupling
that preserves marginal distributions, and showed that
minimum-entropy coupling maximizes throughput.
Because inverse-transform sampling with any seed produces samples
from the model's true distribution, our PRNG channel satisfies this
perfect security condition by construction.
Ding \emph{et~al.}~\cite{DBLP:conf/sp/DingCWZZY23} (Discop) rotate
probability intervals to create ``distribution copies'' selected by
secret bits, using the shared PRNG seed explicitly as the
steganographic key.
More recently, SparSamp~\cite{DBLP:conf/uss/WangPCDPPH025} and
Shimmer~\cite{DBLP:conf/uss/BaiPLY025} likewise rely on a shared
PRNG seed as the principal secret.

Beyond protocol-level designs, recent work has explored emergent
steganographic risks in deployed LLMs.
Mathew \emph{et~al.}~\cite{DBLP:conf/ijcnlp/MathewMMVWCS25} showed that steganographic
collusion can arise unintentionally in multi-agent systems through
reward misspecification, and Meier \emph{et~al.}~\cite{DBLP:conf/emnlp/MeierWRRG25}
demonstrated that fine-tuned models can covertly exfiltrate context
via steganography at inference time without user awareness.

\bheading{Seed recovery in generative models}
Most closely related to our work, M\"{a}chtle \emph{et~al.}~\cite{DBLP:journals/corr/abs-2509-09488}
demonstrate an analogous seed-recovery
attack against image diffusion models. 
Their tool, SeedSnitch, searches the seed space by comparing the latent noise vector implied by each candidate seed against the encoder's latent
representation of the target image.  The recovered seed is then used
to conduct a prompt-stealing attack.
While M\"{a}chtle et~al.\ recover seeds in order to \emph{steal
prompts} from diffusion-generated images, we recover seeds from
LLM-generated \emph{text} to enable covert communication. Thus, we differ in model type, modality and goal. However, 
together, the two works suggest that PRNG-based seed recovery is possible across modalities.

\subsection{Security Assumptions in Provably Secure
Stegosystems}
\label{sec:critique}

\todo[inline]{Nur so eine Idee: müssen wir hier evtl. ein bisschen milder werden? Also wir haben das ja nur für \toppone{} gezeigt? Können wir hier sonst was aus der BA nutzen @Ben?}
\todo[inline,author=BW,color=green!50!white]{In der BA hab ich im Endeffekt hauptsächlich über prompt-stealing argumentiert, dass die Verteilung und der Prompt als bekannt angenommen werden sollten. }
\todo[inline,author=SB,color=blue!50!white]{Ich hab es noch ein bisschen abgeschwächt, zum Beispiel "early steganography paper" und ähnliches, so dass sich niemand auf den Fuß getreten fühlt.}

For provably secure steganography, our findings have the following implication: 
numerous recent stegosystems based on generative models utilize PRNGs for sampling, including Meteor \cite{DBLP:conf/ccs/KaptchukJ0R21}, Discop \cite{DBLP:conf/sp/DingCWZZY23}, SparSamp \cite{DBLP:conf/uss/WangPCDPPH025} and Shimmer \cite{DBLP:conf/uss/BaiPLY025} and the broader framework introduced in \cite{DBLP:conf/uss/LiaoYS025}, using the seed as a shared secret. 
Of these works only Shimmer and Meteor explicitly allow the adversary access to the history of the channel (prompt), while in the others the modeling is either implicit or ambiguous. 
In addition, these works take quite different approaches to proving and modeling security, making direct comparisons between them difficult, despite the similarities between their constructions. 
Our empirical results from the unknown-prompt setting indicate that a secret prompt is likely unable to further protect the secrecy of the seed in a way relevant to typical provable security notions, at least not for the choice of \toppone{}. 
Therefore, our experiments serve as an indicator that the generative model should only be viewed as a mechanism to facilitate the underlying channel and not one that can support the security of message-encoding.
This both supports and expands on the observation made by Kaptchuk \emph{et~al.} in Meteor, who argued that relaxed adversarial models in early steganography papers from NLP-conferences were insufficient for an assessment of provable security. 
In many of these earlier works, adversaries were notably denied access to the exact generative model, while our results indicate that the natural threat model used for comparable schemes may be the one used in Meteor and Shimmer, making the prompt and, in turn, the distribution derived from the generative model explicitly known to the adversary.
As observed in the previous section, the feasibility of seed recovery in diffusion models demonstrated in \cite{DBLP:journals/corr/abs-2509-09488} allows for the same conclusion to be made for schemes like Pulsar \cite{DBLP:conf/ccs/JoisBK24} using diffusion models in combination with a PRNG.

\section{Conclusion and Future Work}

We have demonstrated that deterministic PRNG-based decoding in standard LLM inference stacks constitutes a steganographic channel capable of embedding and recovering secret payloads without any modification to model weights, sampling algorithms, or output distributions. The channel exploits the deterministic mapping from seed to token sequence: a sender encodes a message in the PRNG seed, and a receiver reconstructs token-level probability intervals from the generated text to identify the seed by exhaustive search. Our communication  model is naturally satisfied by two parties that use the common \llamacpp{} inference software.

Extensive experiments across six model families and five text domains validate the channel in two operational modes. In the \emph{known-prompt} setting, forced alignment with proactive branching recovers the full 32-bit seed with up to 100\% accuracy, depending on model and text domain from as few as 300~tokens in approximately 34~seconds on a single GPU. In the \emph{unknown-prompt} setting, where only the generated text is available, approximate interval reconstruction combined with offset-tolerant hit-count scoring achieves near-perfect accuracy at 600--800~tokens in about 12~seconds. Temperature is the dominant factor: higher values narrow probability intervals, increasing per-token discriminative power, while top-$k$ filtering has a smaller effect. Notably, a simple empty prompt suffices for interval reconstruction; attempts to approximate the original prompt offer no benefit and can be counterproductive.

\bheading{Future Work}
Several directions merit further investigation. Extending the channel to larger seed spaces (e.g., 64-bit) would increase payload capacity but requires more efficient search algorithms, such as meet-in-the-middle strategies or lattice-based approaches, to keep recovery tractable. Developing methods that relax the Constant-Draw Assumption would enable the channel under nucleus sampling with $p < 1$, broadening applicability to more deployment configurations. Finally, investigating the channel's robustness under post-processing such as paraphrasing, partial truncation, or token-level perturbations would clarify its practical resilience in adversarial settings.

\begin{credits}
\subsubsection*{\ackname} 
Generative AI was utilized during programming, editing,
and grammar enhancement of this work. This work was supported by the Federal Ministry of Research, Technology and Space (BMFTR) through the \href{https://anomed.de/}{\textit{AnoMed}} project.
\end{credits}

\bibliographystyle{plain}
\bibliography{References}

\end{document}